\documentstyle[twocolumn,aps,pra,epsf]{revtex}

\def\psfig#1#2{ \begin{center}
                   \epsfxsize=#2
                   \leavevmode\epsffile{#1.EPS}
                \end{center} }

\newcommand{\ket}[1]{\left| #1 \right\rangle}

\def\RB{{\large $+$}}
\def\DB{{\large $\times$}}

\begin{document}
\draft

\title{The effect of multi-pair signal states in quantum cryptography
       with entangled photons}

\author{Miloslav Du\v sek$^1$ and Kamil Br\'adler$^2$}

\address{$^1$Department of Optics, Palack\'y University, 17.~listopadu 50,
         772~00 Olomouc, Czech~Republic}

\address{$^2$Department of Chemical Physics and Optics, Charles University,
         Ke~Karlovu 3, 121~16 Prague~2, Czech~Republic}

\date{\today}

\maketitle


\begin{abstract}
Real sources of entangled photon pairs (like parametric down conversion)
are not perfect. They produce quantum states that contain more than only
one photon pair with some probability. In this paper it is discussed
what happens if such states are used for the purpose of quantum key
distribution. It is shown that the presence of ``multi-pair'' signals
(together with low detection efficiencies) causes errors in transmission
even if there is no eavesdropper. Moreover, it is shown that even the
eavesdropping, that draws information only from these ``multi-pair''
signals, increases the error rate. Information, that can be obtained by
an eavesdropper from these signals, is calculated.
\end{abstract}



\section{Introduction}

The only evincibly secure method of communication with guaranteed
privacy is Vernam cipher (or one-time pad) \cite{vernam26a}. It requires
both communicating parties share a secret key of the same length as the
message. Quantum key distribution (QKD) is a technique to provide two
parties with such a secure, secret and shared key. The first complete
protocol for QKD was given by Bennett and Brassard \cite{bennett84a}
(BB84) following Wiesner's ideas \cite{wiesner83a}. The essence
of this protocol is that if non-orthogonal quantum states are used for
communication and a channel transmit them perfectly then eavesdropping
is detectable. Later a different protocol, inspired by Bell's
inequalities, was proposed by Ekert \cite{Ekert}. It relies on
nonclassical correlations or entanglement of two quantum particles. Its
simplified (``BB84-like'') version works as follows: Let us suppose two
communicating parties, {\em Alice\/} and {\em Bob\/}, share a set of
entangled pairs $(\ket{V}_A \ket{V}_B+\ket{H}_A \ket{H}_B)/\sqrt{2}$,
where $\ket{V}$ and $\ket{H}$ are two orthonormal states of each
particle -- e.g., vertical and horizontal linear polarizations of
photons. Alice and Bob choose randomly and independently between two
conjugated polarization bases -- e.g. between basis $\{V,H\}$ (``\RB'')
and the ``diagonal'' basis (``\DB'') rotated by $45^\circ$ with respect
to it. Following a public discussion about the basis of the measurement
apparatuses, Alice and Bob can obtain a shared key made up from those
signals where the measurement devices give correlated results. This is
so called sifted key.

Photon pairs with correlated polarizations can be prepared, e.g., by
parametric down conversion of type II \cite{typII} or using two
down-conversion crystals with phase matching of type I \cite{typI}.
Unfortunately, these techniques never produce exactly one pair of
photons. Quantum states generated by the both above mentioned
down-conversion methods should be the same in principle. However, the
system with two non-linear crystals is perhaps
more graphical for our purposes.
The orientation of the optical axes of
two identical crystals are mutually perpendicular. With a vertically
(horizontally) polarized pump beam down-conversion will only occur in the
first (second) crystal, respectively.  A $45^\circ$-polarized pump
photon will be equally likely to down-convert in either crystal.
Let us suppose two spatial modes with two fixed frequencies fulfilling
phase-matching conditions. One is aiming to Alice, the other to Bob. The
first crystal generates beams with vertical polarizations, the second
one with horizontal polarizations. Quantum state generated by one
crystal can be described \cite{WallsMilb}
as\footnote{Of course, this is just an approximation because more then
     only two modes are always present in real cases. If the number of
     signal or idler modes is effectively infinite then the total number
     of photons in signal or idler beam, respectively, obeys Poissonian
     statistics.}
 \begin{equation}
   \ket{\psi} = \xi \sum_{n=0}^{\infty} g^n \ket{n}_A \ket{n}_B,
 \label{stav1}
 \end{equation}
where $\ket{n}$ are corresponding number states, $\xi=(\cosh \chi
t)^{-1} = \sqrt{1-g^2}$, and $g=\tanh \chi t$ with $\chi$ being
proportional to non-linear susceptibility and pump power and $t$ being
interaction
time.\footnote{It has good physical meaning only for pulse-pumped down
     conversion. Then it may be limited to infinity.}
The total quantum state originating from the both crystals is
then\footnote{We neglect a slight decrease of the pump power behind the
     first crystal.}
 \begin{eqnarray}
   \ket{\Psi} & = & \ket{\psi}_1 \ket{\psi}_2 \nonumber \\
   & = & \xi^2 \sum_{m=0}^{\infty} \sum_{n=0}^{\infty} g^{m+n}
   \ket{m}_{AV} \ket{m}_{BV} \ket{n}_{AH} \ket{n}_{BH},
 \label{stav2}
 \end{eqnarray}
where the subscripts $V$ and $H$ denote modes with vertical polarization
(produced by the first crystal) and horizontal polarization (coming
from the second crystal), respectively. The mean number of pairs is
 \begin{equation}
   \mu = \xi^4 \sum_{m,n} (m+n) g^{2(m+n)} = {2 g^2 \over 1-g^2}.
 \label{mu}
 \end{equation}

The presence of more than one pair (or more than one photon in
``single-photon'' protocols) in the signals may enable eavesdropper
({\em Eve\/}) to get some information on the cryptographic key without
causing any error. Thus she can learn something about the key but stay
undisclosed. Similar difficulties implied by the use of weak coherent
states in combination with lossy lines has been pointed out earlier
\cite{huttner95a,yuen96a,dusek99a,nl00a}. A comprehensive analysis of
security aspect of practical quantum cryptosystems taking into account
the source imperfections were done in Ref.~\cite{BrassLut}. But the role
of down-conversion sources was reduced just to the preparation of approximate
single photon states there.
In the present paper we want to go beyond this limitation.

The article is organized as follows.
In Sec.~\ref{SimplErr} we explain,
on a simplified signal state containing at most two pairs of photons, why
errors appear in QKD. Imperfect detection efficiency and losses
on the line are taken into account.
Sec.~\ref{compar} contains the comparison of the amount of information
that can be obtained by Eve from multi-pair or multi-particle signals
(by means of photon-number-splitting attack \cite{BrassLut}) for
different cryptographic schemes. Particularly for quantum cryptography
using entangled photons, weak coherent states, and down-conversion
``single-photon'' sources. In Sec.~\ref{multiclicks} we briefly discuss
restrictions on Eve's activity stemming from monitoring both the data
rate and the ``double-click'' rate (both detectors corresponding to
logical 1 and 0 fire together).
Sec.~\ref{concl} concludes the article with a short summary.

\begin{figure}
\psfig{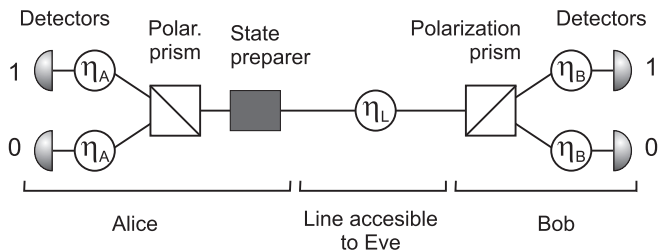}{\hsize}
\caption{Arrangement for QKD. State preparer, situated at Alice's side,
  generates signal states (\protect\ref{stav2}). Both Alice and Bob have
  detectors that cannot distinguish the number of impinging photons and
  whose detection efficiencies are $\eta_A$ and $\eta_B$, respectively
  (this is indicated by circles in the figure). Alice and Bob change
  between two orientations of their polarization analyzers: \protect\RB\
  and \protect\DB. Both communicating parties are connected by quantum
  channel with transmittance $\eta_L$. This channel is accessible to
  Eve.}
\label{fig1}
\end{figure}

\section{Errors in QKD due to imperfect signal states}
  \label{SimplErr}

Consider configuration for QKD as in Fig.~\ref{fig1}. Let us suppose
that $g \ll 1$ so that in Eq.~(\ref{stav2}) we can neglect all terms
containing more than two pairs:
 \begin{eqnarray}
   &&\ket{\Psi}  = \xi \big[
        \ket{0,0,0,0} +
   g    \big( \ket{0,0,1,1} + \ket{1,1,0,0} \big)
        \nonumber \\ &&~~ +\,
   g^2  \big( \ket{0,0,2,2} + \ket{2,2,0,0} +
        \ket{1,1,1,1} \big) + {\cal O}(g^3)
   \big].
 \label{2pairs}
 \end{eqnarray}
Here we have used notation
 \begin{eqnarray}
   |m,m,n,n&& \rangle =
   \ket{m}_{AV} \ket{m}_{BV} \ket{n}_{AH} \ket{n}_{BH}
   \nonumber \\ &&= {1 \over m! n!} \left[
   \left( {\sf a}^{\dag}_{AV}
          {\sf a}^{\dag}_{BV} \right)^m
   \left( {\sf a}^{\dag}_{AH}
          {\sf a}^{\dag}_{BH} \right)^n \right]
   \ket{\rm vac}
 \label{notation}
 \end{eqnarray}
with ${\sf a}^{\dag}$ being creation operators in corresponding modes.

In the diagonal basis
\DB, represented by the following creation operators
 \begin{equation}
  \begin{array}{rcl}
    {\sf a}^{\dag}_X &=& ( {\sf a}^{\dag}_V
                             + {\sf a}^{\dag}_H ) / \sqrt{2}, \\
    {\sf a}^{\dag}_Y &=& ( {\sf a}^{\dag}_V
                             - {\sf a}^{\dag}_H ) / \sqrt{2},
  \end{array}
 \label{Rbase}
 \end{equation}
state (\ref{2pairs}) does {\em not change\/} its form.
It can be shown that even the full  state (\ref{stav2}) is invariant
under such transformations of bases (the same transformation at the both
sides).

Losses on the channel and non-perfect efficiency of Alice's and Bob's
detectors are modeled by beam splitters with intensity transmittances
$\eta_L$, $\eta_A$, and $\eta_B$, respectively.
All the detectors are assumed to be ``yes/no'' detectors, which either
fire or do not fire -- they cannot distinguish the number of
impinging photons. They may be described by the pair of projectors:
${\sf P}_{\rm no}=| 0 \rangle \langle 0 | + \sum_{n=1}^\infty
(1-\eta)^n | n \rangle \langle n |$
and ${\sf P}_{\rm
yes}=\sum_{n=1}^\infty [1-(1-\eta)^n] | n \rangle \langle n |$, where
$\eta$ is a detector efficiency. We neglect any noise.

We intend to show that if the detector efficiencies are lower than
100\,\% the use of signal states
(\ref{2pairs}) causes errors in the sifted key inevitably. Therefore we
are interested only in that cases when Alice and Bob have set the same
polarization bases. Of course, Alice and Bob include to the key only that
events when {\em exactly one\/} detector fires at each side.
The average relative length of the sifted key
(with respect to the number of all generated entangled states) is then
given by the
formula\footnote{It is taken into account that only one half of Alice's
                 and Bob's bases coincide in average.}
\begin{eqnarray}
  R_{\rm key} &\approx & \xi^2 g^2 \big\{
  \eta_A \eta_B \eta_L
  \nonumber \\  &+&
  g^2 \left[ 1 - (1 - \eta_A)^2 \right]
  \left[ 1 - (1 - \eta_B \eta_L)^2 \right]
  \nonumber \\  &+&
  2 g^2 \eta_A (1-\eta_A)\, \eta_B \eta_L (1-\eta_B \eta_L)
  \big\}.
 \label{key}
\end{eqnarray}
On the other hand, the relative number of errors (i.e. events when Alice
gets a bit different from that detected by Bob) is
\begin{equation}
  R_{\rm err} \approx \xi^2 g^4
  \eta_A (1-\eta_A)\, \eta_B \eta_L (1-\eta_B \eta_L).
 \label{err}
\end{equation}
Thus the error rate reads
\begin{eqnarray}
    \varepsilon &=& {R_{\rm err} \over R_{\rm key}} \approx
    { g^2 (1 - \eta_A - \eta_B \eta_L + \eta_A \eta_B \eta_L)
    \over
    1 + g^2 (6 - 4\eta_A - 4\eta_B \eta_L + 3\eta_A \eta_B \eta_L)}
    \nonumber \\ & = &
    { (1-\eta_A) (1-\eta_B \eta_L) \over 2} \mu + {\cal O}(\mu^2).
 \label{ErrRate}
\end{eqnarray}
Clearly, if $\eta_A \to 1$ then $\varepsilon \to 0$ for all mean pair
numbers $\mu$. So Alice should have
as good detectors as possible. At Bob's side the crucial limitation
would probably be represented by a low line transmission $\eta_L$ for real
systems. If $\eta_L \ll \eta_A , \eta_B$ then $\varepsilon \approx
(1-\eta_A)\, \mu /2$.

\section{Information leaked to Eve}
  \label{compar}

Let us suppose now that Eve will try to get some information on the key
only from ``multi-particle'' (or ``multi-pair'') signals in order not to
make any errors in transmission. She will be allowed to use the most
efficient individual attack of this kind -- the photon-number-splitting
(PNS) attack \cite{BrassLut}: She substitute a lossy line by a lossless
one. Then she measures the total number of photons in incoming signals.
If this number is higher than one she extracts and store one photon (or
more). The rest is sent to Bob by her. It is also supposed that she can
control Bob's detection efficiency, so that Bob always get it. If the
number of incoming photons is equal to one she either blocks the signal
or passes it without other changes to Bob (in order not to decrease the
data rate). After the public comparison of Alice's and Bob's bases she
makes a polarization measurement on the stored photons.

The average Eve's information about sifted-key bits is
 \begin{equation}
    I_E = \sum_i r_i \left[ 1 + p_i \log_2 p_i + (1-p_i) \log_2 (1-p_i)
          \right],
 \label{IE}
 \end{equation}
where $r_i$ is a portion of bits that Eve knows with probability
$p_i$; $\sum_i r_i =1$. If Eve knows $r$ per cent bits for certain and
she has no idea about the others then simply $I_E = r$.

\subsection{Weak coherent states}

First let us look at the case of quantum cryptography with weak coherent
states (WCS). The expected average relative length of the sifted key (in
proportion to the number of all sent signals) is \cite{dusek99a,BrassLut}
$$
  R_{\rm exp} = {1 \over 2} \left[1 - \exp(-\eta_L \eta_B \mu') \right],
$$
where $\mu'$ is a mean photon number in a signal state, $\eta_B$ denotes
Bob's detector efficiency. The average
relative number of ``multi-photon'' signals is given by the formula
$$
  R_{\rm multi} = {1 \over 2} \left[1 - (1+\mu') \exp(-\mu') \right].
$$
Eve can learn all the bits stemming from these ``multi-photon'' signals
with certainty.
Thus the information leaked to Eve reads
 \begin{equation}
   I_E^{(\rm WCP)} = \left\{
   \begin{array}{cr}
        1 & \mbox{if~~} R_{\rm exp} \le R_{\rm multi}, \\ \\
        \displaystyle
        {R_{\rm multi} \over R_{\rm exp}} & \approx
        \displaystyle
          {1 \over 2 \eta_L \eta_B} \mu',
        ~\mbox{otherwise}.
   \end{array} \right.
 \label{I-WCP}
 \end{equation}

\subsection{Parametric down conversion}

Now, what information may leak to Eve if a parametric down-conversion
(PDC) source of ``single'' photons is used instead of laser producing
coherent states? Generated signal states (with fixed polarizations) are
used for BB84 QKD-protocol in the exactly same manner as WCS
\cite{BrassLut}. The source consist of a single down-conversion crystal
generating state (\ref{stav1}) and a ``yes/no'' detector (with an
efficiency $\eta_A$) placed in one of the two output modes. A click on this
detector means that the signal state has been prepared at the other
mode.
The expected average relative length of the sifted key (in proportion to
the number of all generated entangled states) is given by the formula
$$
  R_{\rm exp} = {\xi^2 \over 2} \sum_{n=0}^\infty g^{2n}
  \left[ 1- (1-\eta_A)^n \right] \left[ 1 - (1-\eta_L\eta_B)^n \right].
$$
The average
relative number of ``multi-photon'' signals reads
$$
  R_{\rm multi} = {\xi^2 \over 2} \sum_{n=2}^\infty g^{2n}
  \left[ 1 - (1-\eta_A)^n \right].
$$
Again, Eve can learn all the bits carried by the ``multi-photon'' signals
with certainty. After some straightforward
calculations one can find the amount of information leaked to
her:\footnote{It can be done exactly but for our purposes the shown
              approximation is good enough.}
 \begin{equation}
   I_E^{(\rm PDC)} = \left\{
   \begin{array}{cr}
        1 & \mbox{if~~} R_{\rm exp} \le R_{\rm multi}, \\ \\
        \displaystyle
        {R_{\rm multi} \over R_{\rm exp}} & \approx
        \displaystyle
         { 2 - \eta_A \over \eta_L \eta_B}\mu'',
        ~\mbox{otherwise},
   \end{array} \right.
 \label{I-PDC}
 \end{equation}
were we have used the fact that in the case under consideration the mean
number of pairs in each generated entangled state is
$\mu'' = g^2 / (1-g^2)$.

\subsection{Entangled photons}

Finally let us look at the cryptographic scheme fully based on
entanglement of photon polarizations (EP); see Fig.~\ref{fig1}. Signal
states are described by Eq.~(\ref{stav2}). All the detectors are
``yes/no'' ones again; on Alice's side they have efficiencies $\eta_A$,
on Bob's side $\eta_B$.

Here the situation is more complex. It becomes important how many
photons Eve separates. However, we will confine ourselves only to the
simplified situation when at most to pairs are present with a reasonable
probability [see Eq.~(\ref{2pairs})]. Then Eve can separate no more than
one photon and send remaining one to Bob. In contrast to the both cases
described above, now the information $I_{AE}$ that Eve shares with Alice
is {\em different\/} from the information $I_{EB}$ that she shares with
Bob. This is connected with the occurrence of errors in transmission.

The expected rate of sifted-key generation is given by Eq.~(\ref{key}):
$R_{\rm exp}=R_{\rm key}$.
A portion of two-photon signals leaving Alice's terminal -- that signals
that can be read by Eve applying PNS attack -- is
$$
  R_{\rm double} = \xi^2 g^4 \left\{
  \left[ 1 - (1 - \eta_A)^2 \right] +
  \eta_A (1 - \eta_A) \right\}.
$$
The first term represents contributions from states $\ket{0,0,2,2}$
and $\ket{2,2,0,0}$ the second one from $\ket{1,1,1,1}$. Only when
exactly one detector clicks the bit is accepted to the key.

Calculating information it must be taken into account that now Eve does
not know all measured bits with certainty. She cannot distinguish the
signals stemming from states $\ket{1,1,1,1}$ from the other two-photon
signals. And for these particular signals she hits Alice's bit only
with probability 50\,\% and Bob's bit values are even always opposite to
hers. Thus Eve's average information is
 \begin{equation}
   I_j^{(\rm EP)} \approx \left\{
   \begin{array}{lr}
        f(p_j)
        &\mbox{if~~} R_{\rm exp} \le R_{\rm double}, \\ \\
        \displaystyle
         \frac{R_{\rm double}}{R_{\rm exp}} f(p_j)\!\! & \approx
        \displaystyle
         { 3 - 2 \eta_A \over 2 \eta_L \eta_B}\,f(p_j)\,\mu,
        \mbox{~otherwise},
   \end{array} \right.
 \label{I-EP}
 \end{equation}
where $j=AE,EB$ and $f(p_j)=1 + p_j \log_2 p_j + (1-p_j) \log_2
(1-p_j)$. Probabilities that Eve knows Alice's (or  Bob's) bit,
respectively, are given by the ratio of successful results to all
results:
$$
  p_{AE} = {5 - 3 \eta_A \over 6 - 4 \eta_A},
  \quad
  p_{EB} = {2 - \eta_A \over 3 - 2 \eta_A}.
$$
Clearly, $f(p_{EB}) < f(p_{AE}) < 1$ for $\eta_A < 1$ and then
also $I_{EB} < I_{AE} < 1$.
Unfortunately, the fact that the maximum Eve's information [see
Eq.~(\ref{I-EP})] is lower than unity (if $\eta_A<1$) does not represent
any real advantage because for $R_{\rm exp} \le R_{\rm double}$
information $I_{AE}$ is equal to information shared by Alice and Bob,
$I_{AB}= 1 + \varepsilon' \log_2 \varepsilon' + (1-\varepsilon') \log_2
(1-\varepsilon')$.

Notice the other important feature of PNS eavesdropping in EP systems
which is similar to ``single particle'' attacks: If Eve applies PNS
attack by the way described above, i.e. if she tries to reproduce only
the transmission rate ($R_{\rm exp}$), she {\em increases\/} the error
rate. The reason is that she increases the portion of $\ket{1,1,1,1}$
contributions to the key bits. Clearly, this portion gets the following
value: $R^{(E)}_{\rm err} = \xi^2 g^4 \eta_A (1-\eta_A)/2$. Thus due to
eavesdropping the error rate grows to
 \begin{equation}
   \varepsilon' = \left\{
   \begin{array}{lr}
         \displaystyle
          {R^{(E)}_{\rm err} \over R_{\rm double}}  \approx
         \displaystyle
          {1 - \eta_A \over 6 - 4 \eta_A},
        & \mbox{~~if~~} R_{\rm exp} \le R_{\rm double}, \\ \\
         \displaystyle
          {R^{(E)}_{\rm err} \over R_{\rm exp}} \approx
         \displaystyle
          {1 -\eta_A \over 4 \eta_B \eta_L} \mu.
        & \mbox{~~otherwise},
   \end{array} \right.
 \label{errE}
 \end{equation}
The increase of error rate can help to detect an
eavesdropper what is impossible in the analogous situation (PNS attack)
with WCS and PDC systems.

\section{How to restrict Eve's activity}
  \label{multiclicks}

In the previous section Eve was restricted by the demand to reproduce
transmission rate (average number of sifted-key bits) only. However, it
is not the only quantity which could be monitored by Bob. In all the
mentioned techniques Bob can also measure the double-click rate in that
events when he used a different basis than Alice. In case of EP Bob can
even monitor double-click rate in situations with coincident bases (and,
of course, the error rate). Clearly such Bob's activities pose other
important restrictions to Eve \cite{DJL}. Even more possibilities are
offered by passive arrangement, when Alice and Bob do not change bases
actively (see, e.g., Ref.~\cite{pasiv}).

\section{Conclusions}
  \label{concl}

We discussed the effect of ``multi-pair signals'', that inevitably
appear in any system with a parametric-down-conversion source, on the
security of quantum cryptography. We have shown that there is an
important difference between the quantum-cryptographic setup that uses
such a source just as a ``triggered source of photons'' and that which
employs the entanglement of pairs of generated signals directly for
quantum key distribution. In the latter case there is a still nonzero
error rate even if there is no eavesdropper. This is caused by the joint
effect of the occurrence of ``multi-pair signals'' and of low detection
efficiencies. However, the most important result is that in the
latter setup an individual eavesdropping on ``multi-pair signals''
increases the error rate in the transmission.

\section*{Acknowledgments}

M.\,D.\ acknowledges discussions with Zachary Walton. This work has been
supported under the project LN00A015 and Research Plan CEZ:J14 ``Wave
and Particle Optics'' of the Ministry of Education of the Czech Republic
and the project 19982003012 of the Czech National Security Authority.



\end{document}